\documentclass[11pt]{article}
\usepackage{epsf,amsmath,amssymb,graphicx,scalefnt,rotating,enumitem,fancyhdr}
\numberwithin{equation}{section}
\usepackage[acronym]{glossaries}
\let\oldglsentryshort\glsentryshort
\renewcommand{\glsentryshort}[1]{\abbrev{\oldglsentryshort{#1}}}

\usepackage[final]{showkeys}

\usepackage[noadjust]{cite}
\usepackage{filemod}
\usepackage{framed}
\usepackage{comment}
\usepackage[pdftex, colorlinks=true, linkcolor=black, filecolor=black,
  citecolor=black, urlcolor=black, draft=false, bookmarks,
  bookmarksnumbered=true, plainpages=false,
  linktocpage={true}]{hyperref}
\usepackage[affil-it]{authblk}
\usepackage[usenames,dvipsnames]{color}
\usepackage{tikzsymbols}
\usepackage[capitalize]{cleveref}

\newcommand{\apis}{a_\text{s}}
\newcommand{\apie}{a_\text{e}}
\newcommand{\apiehat}{\hat{a}_\text{e}}
\newcommand{\apigf}{a^\text{\gf}_\text{e}}
\newcommand{\apigfhat}{\hat{a}^\text{\gf}_\text{e}}
\newcommand{\apigfhatstar}{\hat{a}^\text{\gf}_{\text{e},\ast}}

\usepackage{slashed}
\usepackage{dsfont}
\usepackage{braket}

\newif\ifshownewacro
\newcommand{\one}{one}

\newcommand{\three}{three}
\newcommand{\four}{four}

\newcommand{\lmup}{l_{\mu p}}
\newcommand{\lmut}{L_{\mu t}}

\newcommand{\ccf}{C_\text{F}}

\newcommand{\ctr}{T_\text{R}}

\newcommand{\nf}{n_\text{f}}
\newcommand{\dR}{d_\text{R}}

\newcommand{\cca}{C_\text{A}}

\newcommand{\bare}{\text{\abbrev{B}}}
\newcommand{\ren}{\text{\abbrev{R}}}
\newcommand{\citere}[1]{Ref.\,\cite{#1}}
\newcommand{\citeres}[1]{Refs.\,\cite{#1}}

\newcommand{\abbrev}[1]{{\scalefont{.9}#1}}
\newcommand{\EulerGamma}{\gamma_\text{E}}
\newcommand{\ep}{\epsilon}

\newcommand{\dd}{\mathrm{d}}
\newcommand{\deriv}[3]{\frac{\partial\ifthenelse{\equal{#1}{}}{}{^{#1}}%
    #2}{\partial #3\ifthenelse{\equal{#1}{}}{}{^{#1}}}}
\newcommand{\dderiv}[3]{\frac{\dd\ifthenelse{\equal{#1}{}}{}{^{#1}}%
    #2}{\dd #3\ifthenelse{\equal{#1}{}}{}{^{#1}}}}
\newcommand{\order}[1]{\ensuremath{{\cal O}(#1)}}
\newcommand{\nklo}[1]{\abbrev{N$^{#1}$LO}}

\newcommand{\msbar}{\ensuremath{\overline{\text{\abbrev{MS}}}}}

\newcommand{\rhs}{r.h.s.}
\newcommand{\wrt}{w.r.t.}
\newcommand{\gf}{\abbrev{GF}}
\newcommand{\myacrodef}[3]{\newacronym[shortplural=\abbrev{#2}s]{#2}{#2}{#3}\newcommand{#1}{\gls{#2}}}
\newcommand{\QEDthree}{\abbrev{QED}$_3$}
\newcommand{\QEDfour}{\abbrev{QED}$_4$}

\newacronym[plural=effective field theories,shortplural=\abbrev{EFT}s]{EFT}%
{EFT}{effective field theory}

\myacrodef{\QED}{QED}{quantum electrodynamics}
\myacrodef{\rge}{RGE}{renormalization-group equation}
\myacrodef{\ibp}{IBP}{integration-by-parts}
\myacrodef{\qft}{QFT}{quantum field theory}
\myacrodef{\sftx}{SFTX}{short-flow-time expansion}
\myacrodef{\vev}{VEV}{vacuum expectation value}
\myacrodef{\rg}{RG}{renormalization group}
\myacrodef{\gff}{GFF}{gradient-flow formalism}
\myacrodef{\ope}{OPE}{operator product expansion}
\newcommand{\qcd}{\abbrev{QCD}}
\myacrodef{\lhc}{LHC}{Large Hadron Collider}
\myacrodef{\uv}{UV}{ultraviolet}
\myacrodef{\ir}{IR}{infrared}
\myacrodef{\lo}{LO}{leading order}
\myacrodef{\nlo}{NLO}{next-to-leading order}
\myacrodef{\nnlo}{NNLO}{next-to-next-to-leading order}
\myacrodef{\llog}{LL}{leading logarithmic}
\myacrodef{\nll}{NLL}{next-to-leading logarithmic}
\myacrodef{\nnll}{NNLL}{next-to-next-to-leading logarithmic}
\myacrodef{\pdf}{PDF}{parton distribution function}

\myacrodef{\sm}{SM}{Standard Model}
\myacrodef{\bsm}{BSM}{beyond-the-\gls{SM}}
\myacrodef{\mssm}{MSSM}{Minimal Supersymmetric \gls{SM}}
\myacrodef{\susy}{SUSY}{Supersymmetry}
\myacrodef{\dreg}{DREG}{Dimensional Regularization}
\myacrodef{\dred}{DRED}{Dimensional Reduction}
\myacrodef{\ckm}{CKM}{Cabbibo-Kobayashi-Maskawa}
\myacrodef{\hqe}{HQE}{Heavy quark expansion}
\myacrodef{\dis}{DIS}{Deep inelastic scattering}
\myacrodef{\gpd}{GPD}{generalized parton density}
\allowdisplaybreaks
\newcommand{\RHheaderline}{\textsf{TTK-25-44, P3H-26-006, January 2026}}
\fancypagestyle{firstpage}
{
  
  \fancyhead[R]{\RHheaderline}
}
\title{The gradient-flow coupling\\ of three-
  and four-dimensional QED}
\author{Lars Georg}
\author{Robert V. Harlander}
\author{Robert H. Mason}
\affil{TTK, RWTH Aachen University, 52056 Aachen, Germany}

\date{}
\glsresetall 

\begin{document}
\maketitle
\thispagestyle{firstpage}
\begin{abstract}
  We evaluate the \abbrev{QED} coupling in the gradient-flow scheme in three
  and four space-time dimensions. Our general result applies to any theory
  with a U(1) gauge field coupled to arbitary other fields via arbitrary
  interactions. As an example, we consider \abbrev{QED} with $\nf$ flavors in
  three and four space-time dimensions and evaluate the corresponding $\beta$
  functions. In four dimensions, we find that the perturbative expansion of
  the $\beta$ function behaves much better than the corresponding expression
  in \abbrev{QCD}. In three dimensions, we recover both the ultraviolet as
  well as the infrared fixed points of the \abbrev{QED} coupling in the
  large-$\nf$ limit.
\end{abstract}
\parskip.0cm
\tableofcontents
\parskip.2cm
\glsresetall 

\section{Introduction}
\label{sec:intro}

As first-principle descriptions of some central aspects of \qcd\ such as
dynamical chiral symmetry breaking or confinement are still missing, simpler
theories with similar properties may be helpful in gaining insight into
theoretical concepts that can induce such phenomena. A very popular such
theory is \QED\ in three space-time dimension, henceforth referred to as
\QEDthree\ in this paper. Its coupling has positive mass dimension, which
makes the theory super-renormalizable. On the other hand, the theory becomes
strongly interacting at low energies. This prohibits a naive perturbative
approach to \QEDthree. However, the \ir\ behavior is softened when considering
a large number of fermion flavors $\nf$~\cite{Appelquist:1981vg}, in which
case the theory develops an \ir\ fixed
point~\cite{Appelquist:1986fd}. Systematically resumming powers of $1/\nf$
thus allows one to resort to a perturbative approach. A comprehensive review
of large $\nf$ quantum field theory can be found in \citere{Gracey:2018ame}.

A particularly attractive feature of \QEDthree\ is the dynamical generation of
a fermion mass term via spontaneous chiral symmetry
breaking~\cite{Pisarski:1984dj}. However, \QEDthree\ also receives a lot of
attention from condensed matter physics, for example in the context of
high-temperature superconductors (see, e.g., \citere{Franz}).

The resemblence to \qcd, in particular its non-perturbative character,
suggests a lattice approach to \QEDthree~\cite{Dagotto:1988id}. Traditionally,
perturbative and lattice results are difficult to reconcile due to their
different ways of regularizing the divergences of \qft. However, since the
advent of the \gff, this obstacle can be overcome, because it provides a
regularization/renormalization scheme which is accessible both perturbatively
and on the lattice~\cite{Narayanan:2006rf,Luscher:2009eq,Luscher:2010iy,
  Luscher:2011bx,Luscher:2013cpa}.

In \qcd, the \gff\ is an established tool for practical lattice calculations,
allowing for an efficient determination of the overall mass scale in a
particular gauge configuration~\cite{Luscher:2010iy}. It extends the
four-dimensional gauge fields to an auxiliary fifth dimension, the flow time
$t$. The dependence of the flowed fields on this parameter is determined by
the flow equation, which drives the fields from the unflowed gauge
configuration towards the classical solution of the field equations.

One of the main features of the \gff\ is that composite operators of
(renormalized) flowed fields do not require operator renormalization. Taken at
finite flow time $t$, matrix elements of such operators can therefore be
extrapolated to the continuum limit individually. The connection to the
physical limit at $t=0$ can be done with the help of suitable matching
coefficients which can be obtained by a perturbative solution of the flow
equations. Examples for applications of this method are found in
\citeres{Suzuki:2013gza,Makino:2014taa,Taniguchi:2020mgg,
  Black:2023vju,Black:2024iwb,Black:2025gft,Francis:2025rya}.

The \gff\ provides a renormalization condition for the gauge coupling which
can be implemented both perturbatively and on the lattice. In \qcd, this
allows one to derive the strong coupling $\alpha_s(M_Z)$, one of the central
input parameters for perturbative calculations, from low-energy hadronic
observables~\cite{Hasenfratz:2023bok,Wong:2023jvr,Schierholz:2024lge,
  FlavourLatticeAveragingGroupFLAG:2024oxs,Larsen:2025wvg,Brida:2025gii}. It
was found that, while the conversion between the coupling from the \msbar\ to
the \gf\ scheme is perturbatively well behaved, the corresponding $\beta$
function suggests a much smaller radius of convergence compared to its
\msbar\ equivalent~\cite{Harlander:2016vzb,Harlander:2021esn,
  Hasenfratz:2023bok}.

In this paper, we study the analoguous quantity in \QED, both in four and in
three space-time dimensions. We show that, due to the linearity of the
\QED\ flow equations, this coupling is given by the Laplace transform of the
full photon propagator in the unflowed theory. Given the latter, the
\gf\ \QED\ coupling can thus be obtained without any approximation. As an
example, we evaluate this quantity in \four-dimensional \QED\ (\QEDfour),
using the perturbative \QED\ expression for the photon polarization function
through \four-loop level. We compare the resulting $\beta$ function to the
\msbar\ scheme and find that, as opposed to \qcd, their perturbative behavior
is very similar.

We then proceed to \QEDthree, where we adopt the large-$\nf$ limit in order to
derive the \gf\ \QEDthree\ coupling and the corresponding $\beta$
function. Using suitable expansions of the result at short and long flow
times, we recover the Gau\ss{}ian fixed point in the \uv, as well as the
non-trivial \ir\ fixed point. 

\section{Flowed QED}
\label{sec:flowed_qed}

We consider a theory in $D$ Euclidean space-time dimensions with a U(1) gauge
factor and $\nf$ massless fermion flavors. We write this theory as
\begin{equation}\label{eq:flow:jump}
  \begin{aligned}
    \mathcal{L} &= -\frac{1}{4}F_{\mu\nu}F_{\mu\nu}
    + \sum_{i=1}^{\nf}\bar\psi_i i\slashed{D}\psi_i
    + \frac{1}{2\xi}(\partial_\mu A_\mu)^2 + \Delta\mathcal{L}
    \,,
  \end{aligned}
\end{equation}
where
\begin{equation}\label{eq:flow:bely}
  \begin{aligned}
    F_{\mu\nu} &= \partial_\mu A_\nu - \partial_\nu A_\mu
  \end{aligned}
\end{equation}
is the U(1) field-strength tensor, $\psi_i$ is the fermon field of flavor $i$,
and $\xi$ is the U(1) gauge parameter.  The dependence of all fields on the
Euclidean space-time variable $x$ is suppressed here and in the following,
unless required for clarity. The term $\Delta\mathcal{L}$ contains arbitrary
other sectors of the theory to which the fields $A_\mu$, $\bar\psi_i$, and
$\psi_i$ may or may not be coupled. For example, some of the $\psi_i$ could be
massless quarks, and $\Delta\mathcal{L}$ describes their weak and strong
interactions. In the following, the U(1) part of the theory will be most
relevant; we will refer to it as \QED, even though in the \sm\ it would
correspond to the hypercharge interactions.

The flowed \QED\ gauge field $B_\mu(t)$ is defined via the flow equation
\begin{equation}\label{eq:flow:gowd}
  \begin{aligned}
    \partial_t B_\mu(t) &= \Box B_\mu(t) +
    (\kappa-1)\partial_\mu\partial_\nu B_\nu(t)\,,
  \end{aligned}
\end{equation}
and the boundary condition
\begin{equation}\label{eq:flow:fist}
  \begin{aligned}
    B_\mu(t=0) &= A_\mu\,,
  \end{aligned}
\end{equation}
where $\kappa$ is a gauge parameter which drops out of physical observables;
throughout this paper, we will set it to $\kappa=1$.

As opposed to flowed \qcd, the flow equation for the \QED\ vector boson
is linear and can be solved exactly in terms of the unflowed field. 
For our choice $\kappa=1$, the Fourier transform of the flowed field is given by
\begin{equation}\label{eq:flow:ieda}
  \begin{aligned}
    \tilde{B}_\mu(t,p) &= \tilde{A}_\mu(p)\,e^{-tp^2}\,,
  \end{aligned}
\end{equation}
In principle one also has to consider an electron flow equation for a complete
gradient flow treatment of \QED. However, unlike the case of flowed \qcd, one
can trivially express the matrix elements of purely photonic flowed
\QED\ operators in terms of regular \QED\ fields. This can be seen in
particular by focusing on the \vev\ of the flowed \QED\ action density,
obtained by
\begin{equation}\label{eq:flow:dupe}
  \begin{aligned}
    E(t) &= \frac{e_\bare^2}{4}\,
    \int\dd^D x\,\langle B_{\mu\nu}(t,x) B_{\mu\nu}(t,x)\rangle
    =
    \frac{e_\bare^2}{4} \int\frac{\dd^D p}{(2\pi)^D}
    \,e^{-2tp^2}\langle F_{\mu\nu}(p) F_{\mu\nu}(-p)\rangle
    \\&=
    \frac{e^2}{2}(D-1)
    \int\frac{\dd^D p}{(2\pi)^D}\frac{e^{-2tp^2}}{1+\Pi_\ren\left(p\right)}\,,
  \end{aligned}
\end{equation}
with the renormalized electric charge $e$ and the renormalized photon
polarization function $\Pi_\ren$.  In diagrammatical form, \cref{eq:flow:dupe}
looks as follows:\footnote{The Feynman diagram has been produced with \texttt{FeynGame}~\cite{Harlander:2020cyh,Bundgen:2025utt}.}
\begin{equation}\label{eq:flow:army}
  \begin{aligned}
    E(t) &= \raisebox{-2.5em}{\includegraphics[width=.15\textwidth]
      {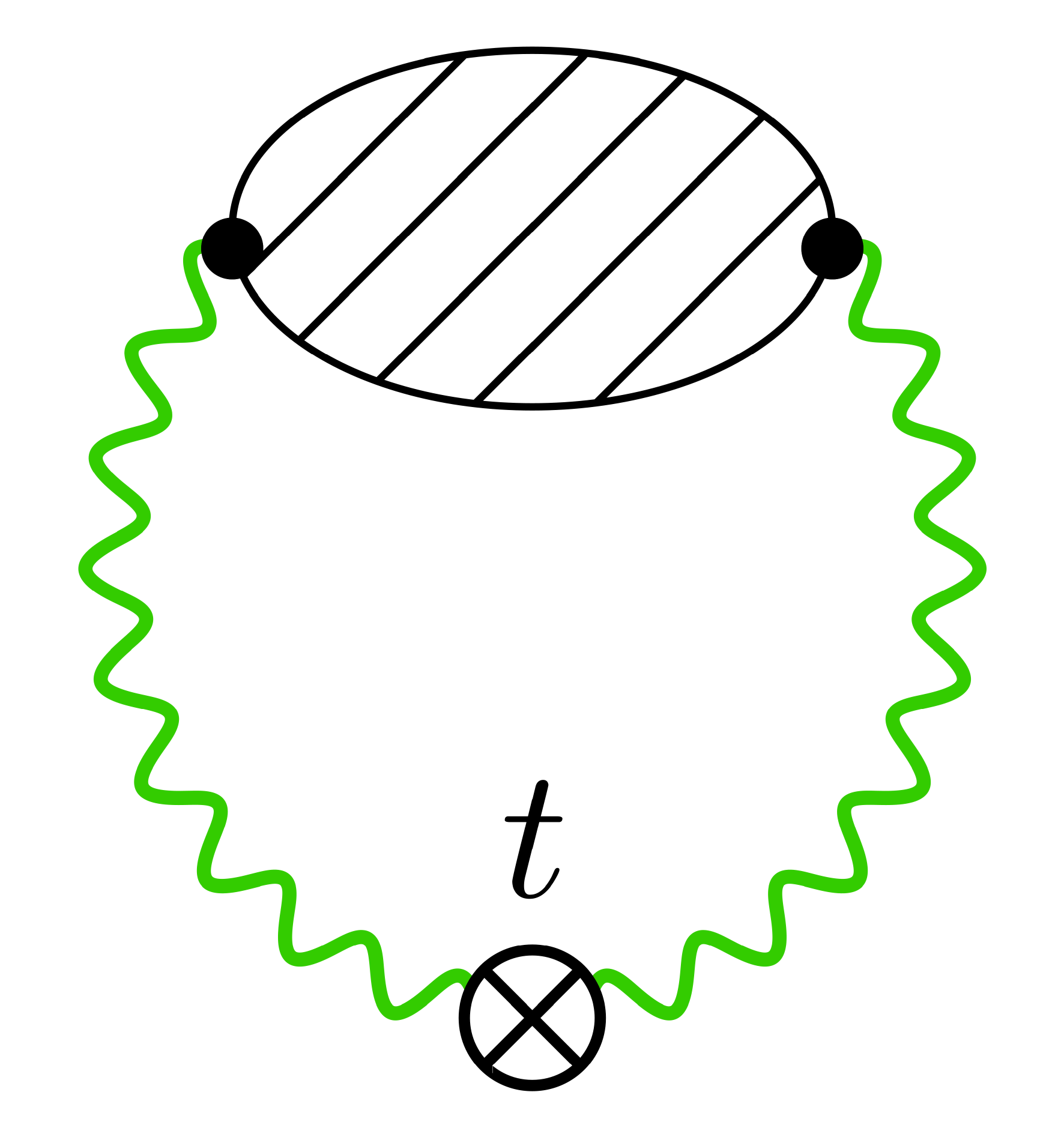}}\,,
  \end{aligned}
\end{equation}
where the shaded blob denotes the regular photon polarization function. It is
important to note that the relation in \cref{eq:flow:army} holds for the full
Lagrangian in \cref{eq:flow:jump}, including the non-\QED\ interactions of
$\Delta\mathcal{L}$. The latter only enter via the polarization function
though.

In the language of the perturbative gradient flow in \qcd, \cref{eq:flow:army}
can be understood as follows: Since the \QED\ flow equation is linear, there
are no outgoing photon \textit{flow lines} from any of the \textit{flowed
  vertices} (cf.~\citere{Artz:2019bpr} for the notation).  Therefore, the
flowed photons $B_\mu(t)$ occuring in the flowed action density must be
connected to the polarization function at vanishing flow time, for which one
can therefore use the expression of the unflowed theory.

Following the definition of the gradient-flow coupling in
\qcd~\cite{Luscher:2013vga}, we may define the \QED\ \gf\ coupling
\begin{equation}\label{eq:flow:bund}
  \begin{aligned}
    \apigf(\mu) &= \frac{(8\pi t)^{D/2}}{2\pi^2(D-1)}
    E(t)\bigg|_{t=(c_{t\mu}/\mu)^2}\,,
  \end{aligned}
\end{equation}
where $c_{t\mu}$ is a fixed, but arbitrary constant, which is required to
fully define the scheme. It may be worth pointing out that, according to
\cref{eq:flow:dupe} and up to a global factor, the gradient-flow coupling in
\QED\ is thus just the Laplace transform of $p^{D-2}/(1+\Pi_\ren(p))$ \wrt\ $p^2$.

Let us stress that \cref{eq:flow:dupe,eq:flow:army} are exact and hold for
arbitrary space-time dimension~$D$ and the general renormalized photon
polarization $\Pi_\ren$. In the following, we will consider $E(t)$ for two
specific cases: (i)~The perturbative approximation in \four-dimensional
\QED\ through \nklo{3} in the \QED\ coupling for general flow time $t$;
(ii)~The large-$\nf$ limit for \three-dimensional \QED\ through \nlo\ in
$1/\nf$ in the limit of short and long flow times. The following integral
will be helpful in this respect:
\begin{equation}\label{eq:flow:blot}
  \begin{aligned}
    I(t,\beta) &= \int\frac{\dd^D p}{(2\pi)^D}\,p^{-\beta} e^{-2tp^2} =
    \frac{(2t)^{\beta/2}}{(8\pi t)^{D/2}}\frac{\Gamma((D-\beta)/2)}{\Gamma(D/2)}\,,
  \end{aligned}
\end{equation}
where $\Gamma(z)$ is Euler's $\Gamma$ function, i.e.\ $z\Gamma(z) =
\Gamma(z+1)$ and $\Gamma(1)=1$. In writing the \rhs\ of \cref{eq:flow:blot},
we assumed that the parameters $D$, $\beta$, and $t$ are such that the
integral exists, which can always be assumed in dimensional regularization.

\section{Four-dimensional QED}\label{sec:4dqed}

As an example for the application of \cref{eq:flow:dupe}, we consider
\QED\ with $\nf$ massless fermions in four space-time dimensions and write the
perturbative expansion of the \msbar-renormalized polarization function as
\begin{equation}\label{eq:pi}
  \Pi_\ren(p)
  =\sum_{n\geq1} c_n(\lmup)\,\apie^n(\mu)
  =\sum_{n\geq1}\sum_{m=0}^nc_{n,m}\,\lmup^m\,\apie^n(\mu)
    \,,
\end{equation}
where
\begin{equation}\label{eq::ilia}
  \begin{aligned}
   \lmup &= \ln\frac{\mu^2}{p^2}\qquad\text{and}\qquad
    \apie(\mu) = \frac{e^2(\mu)}{4\pi^2}\,,
  \end{aligned}
\end{equation}
with the \msbar\ renormalized \QEDfour\ coupling $e(\mu)$ depending on the
renormalization scale $\mu$.  The coefficients $c_{n,m}$ have been obtained
through four loops (i.e.\ $n=4$) by Baikov~et~al.~\cite{Baikov:2012rr}. If we
insert this expression into \cref{eq:flow:dupe} and re-expand in $\apie$, we
arrive at integrals of the form
\begin{equation}\label{eq::jagg}
  \begin{aligned}
    I_m(t)&=\int\frac{\text{d}^Dp}{(2\pi)^D}e^{-2tp^2}\lmup^m
    =\frac{\partial^m}{\partial \beta^m}
    \mu^{2\beta}I(t,2\beta)\bigg|_{\beta=0}\,,
  \end{aligned}
\end{equation}
where one may insert the result of \cref{eq:flow:blot} in order to arrive at
an analytical expression.

\begin{figure}[h]
  \begin{center}
    \begin{tabular}{cc}
      \raisebox{0em}{%
        \mbox{%
          \includegraphics[%
            clip,width=.45\textwidth]%
                          {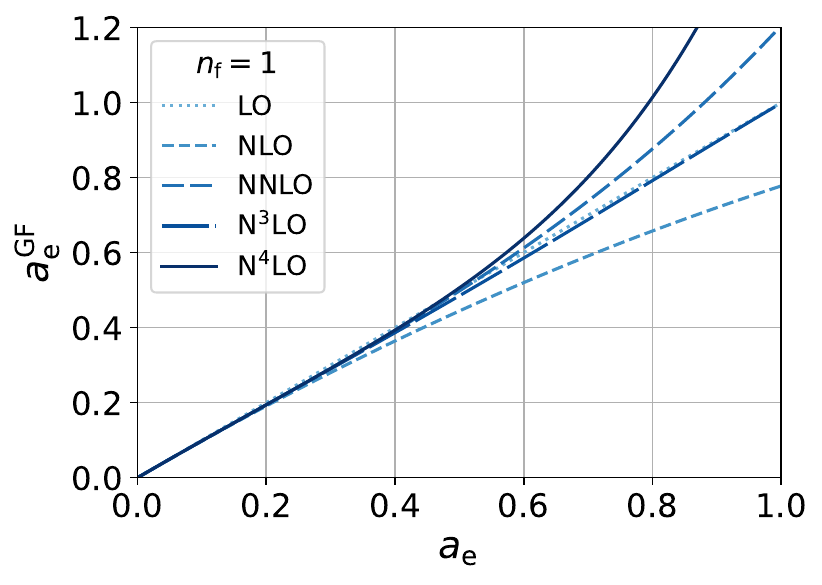}}} &
      \raisebox{0em}{%
        \mbox{%
          \includegraphics[%
            clip,width=.45\textwidth]%
                          {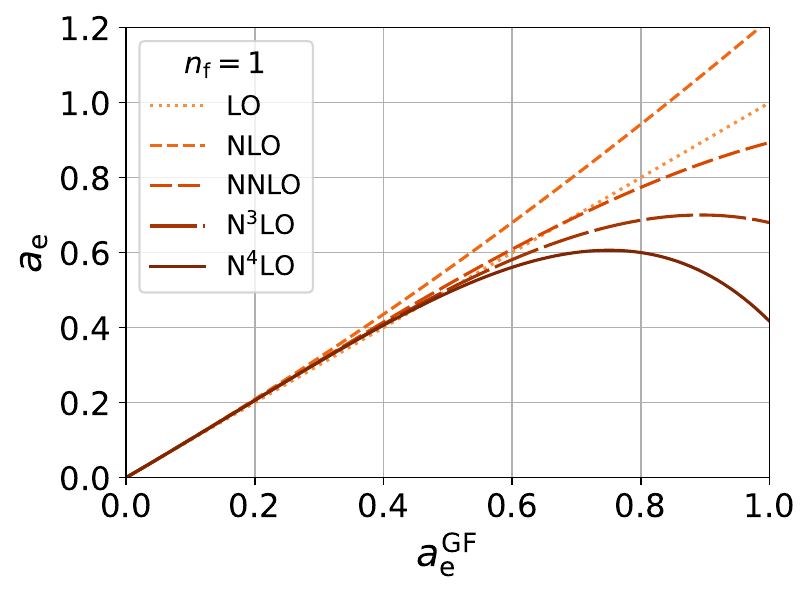}}} \\
    (a) & (b) \\
    \raisebox{0em}{%
        \mbox{%
          \includegraphics[%
            clip,width=.45\textwidth]%
                          {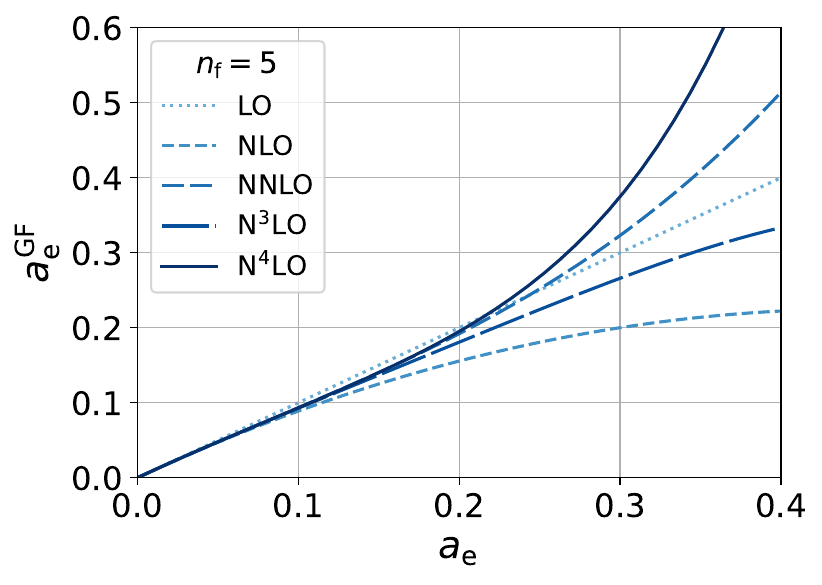}}} &
      \raisebox{0em}{%
        \mbox{%
          \includegraphics[%
            clip,width=.45\textwidth]%
                          {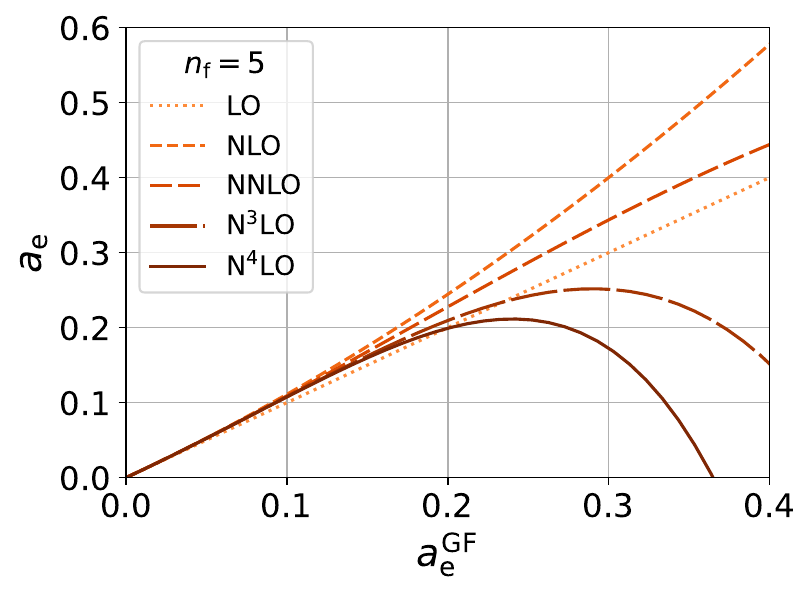}}} \\
      (c) & (d)
    \end{tabular}
    \parbox{.9\textwidth}{
      \caption[]{\label{fig:apiTransfer_n1}\sloppy The  conversion of the coupling from
        the $\msbar$ scheme to the \gf\ scheme (a,c) and
        back (b,d) for \QEDfour. Here we used $c_{t\mu}=e^{-\gamma_E/2}/\sqrt{2}$ and $\nf=1$ (a,b) or $\nf=5$ (c,d).  } }
  \end{center}
\end{figure}

Following \cref{eq:flow:bund}, we obtain
\begin{equation}\label{eq::item}
  \begin{aligned}
    \apigf(\mu) &= \apie(\mu)
    \,\sum_{n\geq 0}\sum_{m=0}^n e_{n,m}\,\apie^n(\mu)
    \,\lmut^m\,,
  \end{aligned}
\end{equation}
where
\begin{equation}\label{eq::berm}
  \begin{aligned}
    \lmut &= \ln 2c^2_{t\mu} + \EulerGamma\,,
  \end{aligned}
\end{equation}
with $\EulerGamma = -\Gamma'(1)=0.577216\ldots$, and the constant $c_{t\mu}$
introduced in \cref{eq:flow:bund}.  The result for the non-logarithmic
coefficients reads
\newcommand\numberthis{\addtocounter{equation}{1}\tag{\theequation}}
\begin{align*}
    e_{0,0} &= 1\,,\qquad e_{1,0} =-\frac{2}{9}\nf\,,\\
    e_{2,0} &=
    \left(\zeta(3)-\frac{43}{48}\right)\nf
    +\left(\frac{\pi^2}{54}-\frac{5}{81}\right)\nf^2\,,\\
    e_{3,0} &= \nf\left[\frac{67}{144}+\frac{37}{24}\zeta(3)
      -\frac{5}{2}\zeta(5)\right]
    +\nf^2\left[\frac{3113}{2592}+\frac{5\pi^2}{144}
     -\frac{7}{6}\zeta(3)\right]\\
   &\quad +\nf^3\left[\frac{100}{729}-\frac{\pi^2}{81}
     -\frac{2}{27}\zeta(3)\right]\,,\\
   e_{4,0}&=\nf\left[-\frac{107}{768}-\frac{13}{32}\zeta(3)
     -\frac{245}{32}\zeta(5)+\frac{35}{4}\zeta(7)\right]\\
   &\quad+\nf^2\left[\frac{299}{512}+\frac{\pi^2}{96}-\frac{11\pi^4}{8640}
     -\frac{97}{12}\zeta(3)+2\zeta^2(3)
     +\frac{115}{18}\zeta(5)\right]\\
   &\quad+\nf^2\ \textbf{si}\left[-\frac{299}{432}+\frac{\pi^4}{360}
     +\frac{31}{48}\zeta(3)+\frac{2}{3}\zeta^2(3)
     -\frac{5}{4}\zeta(5)\right]\\
   &\quad+\nf^3\left[-\frac{26611}{31104}
     -\frac{347\pi^2}{2592}+\frac{47}{216}\zeta(3)+\frac{\pi^2}{9}\zeta(3)
     +\frac{5}{9}\zeta(5)\right]\\
   &\quad+\nf^4\left[-\frac{875}{6561}-\frac{5\pi^2}{729}
     +\frac{\pi^4}{540}+\frac{16}{243}\zeta(3)\right]\,,
   \numberthis\label{eq::fyrd}
\end{align*}
where $\zeta(z)$ is Riemann's $\zeta$ function with values
$\zeta(3)=1.20206\ldots$, $\zeta(5) = 1.03693\ldots$. $\textbf{si}=1$ is a
marker used to separate singlet and non-singlet contributions, analogous to
\citere{Baikov:2012rr}. This conversion between the $\msbar$ and the
$\text{\gf}$ scheme is illustrated in \cref{fig:apiTransfer_n1} for
$\nf=1$ and $\nf=5$.

Instead of quoting the logarithmic coefficients explicitly, we remark that
they can be easily obtained from the \rg\ invariance of the action density,
\begin{equation}\label{eq::jaup}
  \begin{aligned}
    \mu\dderiv{}{}{\mu}E(t) &= 0\,,
  \end{aligned}
\end{equation}
leading to the recurrence relations
\begin{equation}\label{eq::bkcy}
  \begin{aligned}
    e_{0,k} &= 0\,,\qquad
    ke_{n,k}=\sum_{l=0}^{n-1}(n-l)e_{n-l-1,k-1}\beta_{\text{e},l}\,,
    \qquad k\geq 1\,,\ n\geq 1\,,
  \end{aligned}
\end{equation}
with the \QEDfour\ $\beta$ function given by
\begin{align}
    \beta_\text{e}&=-\apie\sum_{k\geq0}\beta_{\text{e},k}\apie^k
\end{align}
where~\cite{Tarasov:1980au,Larin:1993tp,vanRitbergen:1997va,
  Czakon:2004bu,Baikov:2016tgj,Herzog:2017ohr}
\begin{equation}
  \begin{aligned}\label{eq:betae}
    \beta_{\text{e},0}&=-\frac{1}{3}\nf\,,\qquad
    \beta_{\text{e},0}=-\frac{1}{4}\nf\,,\qquad \beta_{\text{e},2}= \frac{11 }{144}\nf^2+\frac{1}{32}\nf\,,\\
    \beta_{\text{e},3}&=\frac{77}{3888}\nf^3+\left(\frac{13}{36}\zeta(3)-\frac{95}{864}\right)\nf^2+\frac{23}{128}\nf\,,\\
    \beta_{\text{e},4}&=-\left(\frac{1}{216}\zeta (3)+\frac{107}{31104}\right)\nf^4+\left(\frac{5}{12}\zeta (5)-\frac{125}{216}\zeta (3)+\frac{13 \pi ^4}{8640}+\frac{10879}{41472}\right)\nf^3\\
    &\quad-\left(\frac{85}{32}\zeta (5)-\frac{31 }{32}\zeta (3)-\frac{3731}{4608}\right)\nf^2-\left(\frac{4157}{6144}+\frac{1}{8}\zeta (3)\right)\nf\,.
  \end{aligned}
\end{equation}
While the conversion from the $\msbar$ to the \gf\ scheme of the coupling
according to \cref{eq::item} depends on the choice of the constant $c_{t\mu}$,
the \gf\ $\beta$ function is independent of it:
\begin{equation}\label{eq::enos}
  \begin{aligned}
    \mu\dderiv{}{}{\mu}\apigf &= \apigf
    \beta_e^\text{\gf}\,,
    \qquad
    \beta_\text{e}^\text{\gf}&=
    -\apigf\sum_{k\geq0}\beta^\text{\gf}_{\text{e},k}\,(\apigf)^k\,.
  \end{aligned}
\end{equation}
The first two coefficients are universal, i.e.
\begin{equation}\label{eq::jane}
  \begin{aligned}
    \beta_\text{e,0}^\text{\gf}
    &= \beta_\text{e,0}\,,\qquad
    \beta_\text{e,1}^\text{\gf}
    &= \beta_\text{e,1}\,,
  \end{aligned}
\end{equation}
see \cref{eq:betae}, while the higher orders are easily obtained from the
results above:
\begin{equation}\label{eq::echo}
  \begin{aligned}
    \beta_\text{e,2}^\text{\gf} &=\beta_{\text{e},2}-\beta_{\text{e},1} e_1+\beta_{\text{e},0}\left(e_2-e_1^2\right)
 \,,\\
 \beta_\text{e,3}^\text{\gf} &=\beta_{\text{e},3}-2 \beta_{\text{e},2} e_1+\beta_{\text{e},1} e_1^2+2 \beta_{\text{e},0} \left(e_3-3 e_2 e_1+ 2 e_1^3\right)\,,\\
 \beta_\text{e,4}^\text{\gf} &=\beta_{\text{e},4}-3\beta_{\text{e},3}e_1-\beta_{\text{e},2}\left(e_2-4 e_1^2\right)+\beta_{\text{e},1}\left(e_3-2 e_1 e_2
\right)\\
&\quad+\beta_{\text{e},0}\left(3 e_4-12 e_3 e_1-5 e_2^2+28 e_2 e_1^2-14
e_1^4\right)\,,
  \end{aligned}
\end{equation}
with
\begin{equation}\label{eq::en}
    e_n=\sum_{m=0}^ne_{n,m}\lmut^m
\end{equation}
and $\lmut$ from \cref{eq::berm}.  The coefficient
$\beta_\text{e,2}^\text{\gf}$ agrees with the \qcd\ result in the Abelian
limit~\cite{Harlander:2016vzb,Harlander:2021esn,
  Hasenfratz:2023bok}; the other coefficients
are new.

\begin{figure}[h]
  \begin{center}
    \begin{tabular}{cc}
      \raisebox{0em}{%
        \mbox{%
          \includegraphics[%
            clip,width=.45\textwidth]%
                          {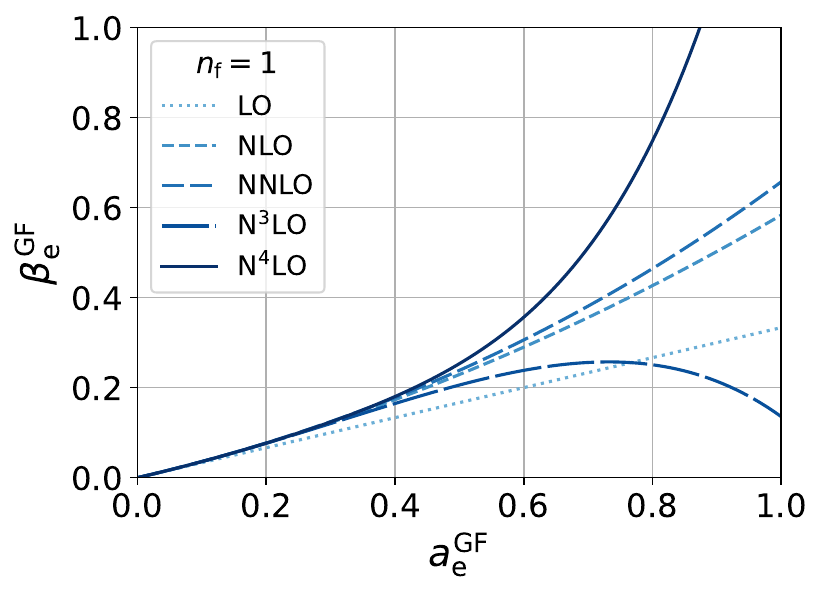}}} &
      \raisebox{0em}{%
        \mbox{%
          \includegraphics[%
            clip,width=.45\textwidth]%
                          {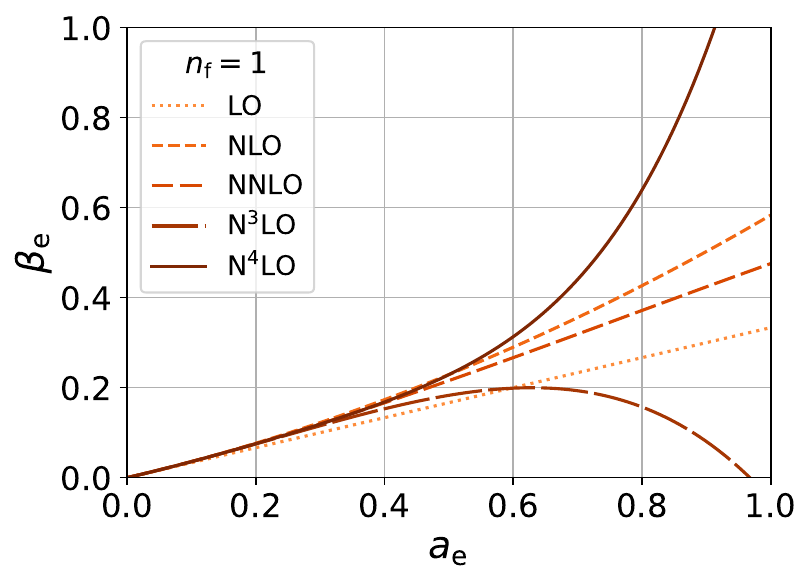}}} \\
      (a) & (b)\\
            \raisebox{0em}{%
        \mbox{%
          \includegraphics[%
            clip,width=.45\textwidth]%
                          {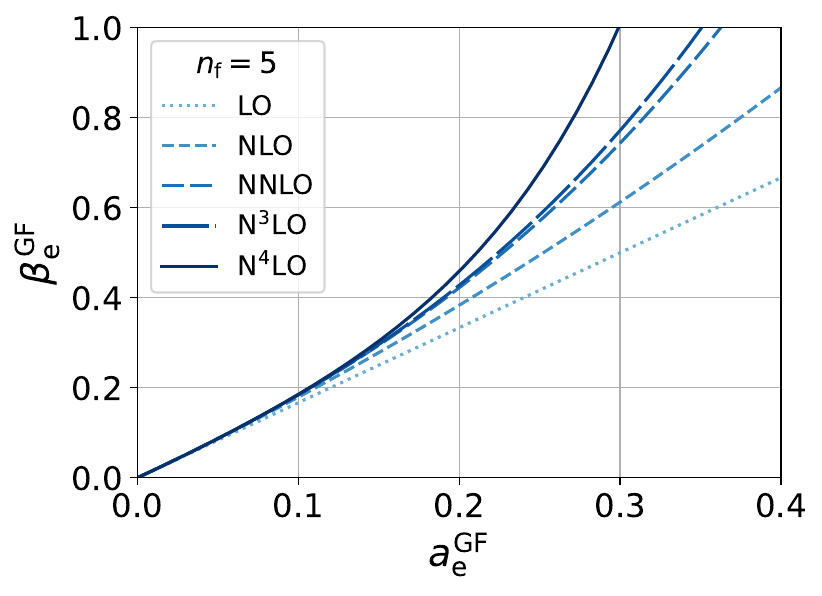}}} &
      \raisebox{0em}{%
        \mbox{%
          \includegraphics[%
            clip,width=.45\textwidth]%
                          {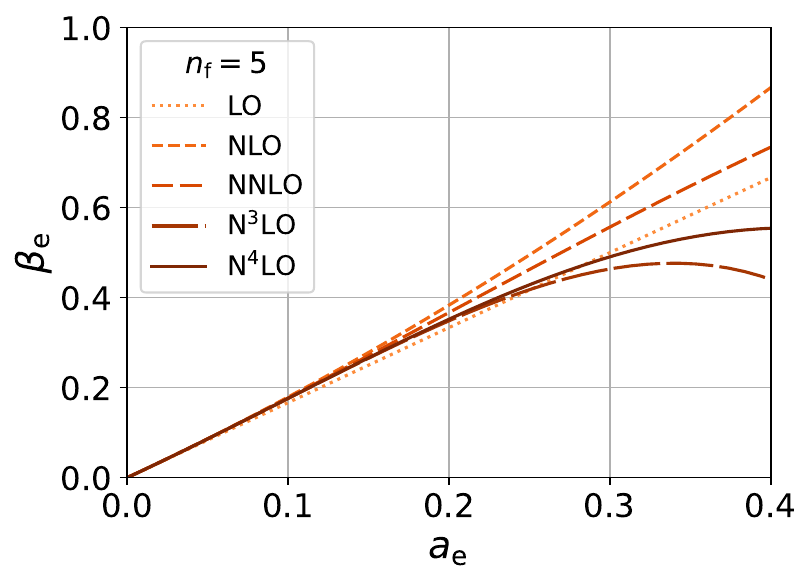}}} \\
      (c) & (d)
    \end{tabular}
    \parbox{.9\textwidth}{
      \caption[]{\label{fig:betaQED4}\sloppy Comparison of the $\beta$ function in the $\text{GF}$ scheme (a,c) to the $\msbar$ scheme (b,d) for $\nf=1$ (a,b) and $\nf=5$ (c,d).  } }
  \end{center}
\end{figure}

The behaviour of the different orders of the $\text{GF}$ $\beta$ function,
together with the $\msbar$ $\beta$ function for $\nf = 1$ and $\nf = 5$ for
comparison, is shown in \cref{fig:betaQED4}. As opposed to the case of
the \qcd\ $\beta$ function, the perturbative behavior in the \gf\ scheme is
rather similar, if not better than in the \msbar\
scheme~\cite{Harlander:2016vzb,Harlander:2021esn,Hasenfratz:2023bok}.

Let us stress that the results of this section should only be considered as an
example for the perturbative application of \cref{eq:flow:dupe}. For example,
one could include the effect of other interactions in the photon polarization
function. In this case, the $c_{n}$ in \cref{eq:pi} become functions of the
coupling constants of these additional interactions. For example, one may
include \qcd\ corrections in the massless-quark limit, which are currently
known through order $\alpha_s^4$~\cite{Baikov:2012zm}. Neglecting higher
orders in $\apie$, the polarization function would then take the form
\begin{equation}\label{eq::alms}
  \begin{aligned}
    \Pi_\ren(p)=\apie(\mu)\,c_{1}(\apis,\lmup)
    = \apie(\mu)\left[\Pi^{(1)} + \sum_{n\geq 1}\sum_{m=0}^n d_{n,m}\,\lmup^m\,\apis^n(\mu)\right]\,,
  \end{aligned}
\end{equation}
and therefore
\begin{equation}\label{eq::does}
  \begin{aligned}
    \apigf(\mu) &= \apie(\mu) (8\pi t)^2\int\frac{\dd^4 p}{(2\pi)^4}
    e^{-2tp^2}\left[1-\Pi_\ren(p)\right]\bigg|_{t=(c_{t\mu}/\mu)^2} +
    \order{\apie^3}\,.
  \end{aligned}
\end{equation}
Here $\Pi^{(1)}$ denotes the \one-loop result, which is of course identical to the $O(\apie)$ term in \cref{eq:pi}, 
i.e.\ $\Pi^{(1)} = c_{1,0} + c_{1,1}\lmup$. Inserting the coefficients $d_{n,m}$ from the literature~\cite{Baikov:2012zm}
and using again \cref{eq::jagg}, one arrives at the corresponding perturbative
expansion for $\apigf$ given in \cref{app:qcd}.

\section{Three-dimensional QED}\label{sec:QED3}

Now we turn to \QED\ in three space-time dimensions. In this case, the
coupling constant $\apie$ has mass dimension one. The theory is
super-renormalizable and strongly coupled in the \ir.

\subsection{Naive perturbation theory}\label{sec:naive}

Despite the non-perturbative \ir\ behavior of \QEDthree, let us consider a
naive perturbative approach for defining the gradient-flow coupling. We will
see that this indeed works up to second order in the coupling, while it fails
beyond that.

The renormalized polarization function is known for general number of
space-time dimensions $D$ through two loops, with the
result~\cite{Grozin:2005yg}
\begin{align}
    \Pi_\ren(q)&=\apie\Pi^{(1)}(q)+\apie^2\Pi^{(2)}(q) + \cdots\,,
\end{align}
with
\begin{align}
  \Pi^{(1)}(q)=\frac{1}{1-D}
  \left(\frac{4\pi}{q^2}\right)^{\frac{4-D}{2}} \frac{D-2}{2} G(1,1)
\end{align}
and 
\begin{align}
    \Pi^{(2)}(q)&=\frac{1}{4(1-D)}
    \left(\frac{4\pi}{q^2}\right)^{4-D}\frac{4D-8}{8-2D}
    \left\{G^2(1,1)\left[-4+\frac{7}{4}D-\frac{1}{4}D^2\right]\right.\\
    &\hspace{2em}\left.+G(1,1)G\left(1,\frac{4-D}{2}\right)
    \frac{4}{4-D}\left[6-5D+\frac{7}{4}D^2-\frac{1}{4}D^3\right]\right\}\,,
\end{align}
where 
\begin{align}
    G(n_1,n_2)&=(4\pi)^{D/2} (p^2)^{n_1+n_2-D/2}
    \int_k\frac{1}{((p+k)^2)^{n_1}(k^2)^{n_2}}\\ &=\frac{\Gamma(n_1+n_2-D/2)
      \Gamma(D/2-n_1)\Gamma(D/2-n_2)}{\Gamma(n_1)\Gamma(n_2)\Gamma(D-n_1-n_2)}\,.
\end{align}
At $D=3$, this gives
\begin{align}
    \Pi^{(1)}(q)=\frac{\nf}{8q}\qquad \textrm{and} \qquad
    \Pi^{(2)}(q)=\frac{\nf}{16\pi^2q^2}\left(\pi^2-10\right)\,.
\end{align}
Inserting this into \cref{eq:flow:dupe} with $D=3$, one may be tempted to
define the \QEDthree\ coupling in the \gf\ scheme as
\begin{equation}\label{eq:QED:bain}
  \begin{aligned}
    \apigf(\mu) &= \apie(\mu)
      \left(1+ \apie(\mu)\sqrt{t}e_1+\apie^2(\mu)te_2 + \cdots
      \right)_{t=(c_{t\mu}/\mu)^2}\,,
  \end{aligned}
\end{equation}
with 
\begin{equation}
    e_1=-\nf\,\pi^2\sqrt{\frac{2}{\pi}}\,,\qquad
    e_2=\nf\pi^2\left(\nf\pi^2-4\pi^2+40\right)\,,
\end{equation}
and $c_{t\mu}$ again arbitrary, but fix.  However, these first two orders are
exceptional in the sense that all higher orders beyond that are
\ir\ divergent, as can be seen as follows. For dimensional reasons, it is
\begin{equation}\label{eq:QED:krti}
  \begin{aligned}
    \Pi^{(n)}(q) \sim 1/q^n\,,
  \end{aligned}
\end{equation}
and thus \cref{eq:flow:dupe} results in integrals of the form
\cref{eq:flow:blot} with $\beta=n\in\mathds{N}$ which do not exist for $n\geq
D=3$. This is a consequence of the well-known property of \QEDthree\ that
$\apie$ is not a viable expansion parameter.

\subsection[Large-$\nf$ limit]{Large-\boldmath{$\nf$} limit}\label{sec:largenf}

Instead, one may consider \QEDthree\ in the limit of large $\nf$. Resumming the
dominant terms in $1/\nf$ leads to the following photon polarization function:
\begin{equation}\label{eq:flow:face}
  \begin{aligned}
    \Pi_\ren(q) &= \frac{\apie}{q}h_f
    \,,
  \end{aligned}
\end{equation}
where~\cite{Gusynin:2000zb,Teber:2012de,Kotikov:2013kcl,Metayer:2023lay}
\begin{equation}\label{eq:QED:ashy}
  \begin{aligned}
    h_f &= \frac{\pi^2\nf}{2}\left(1 + \frac{c^{(1)}}{\pi^2\nf}
    + \order{1/\nf^2}\right)\,,
    \quad\text{with}\quad
    c^{(1)} = \frac{184}{9}-2\pi^2\,.
  \end{aligned}
\end{equation}
Let us define the dimensionless couplings as
\begin{equation}\label{eq:QED:base}
  \begin{aligned}
    \apigfhat(\mu) &\equiv
    c_{\alpha\mu}\frac{\apigf}{\mu}\,,\qquad
    \apiehat(\mu) \equiv
    c_{\alpha\mu}\frac{\apie}{\mu}\,,
  \end{aligned}
\end{equation}
with some constant $c_{\alpha\mu}$. In the following, we choose
\begin{equation}\label{eq:QED:feod}
  \begin{aligned}
    c_{\alpha\mu} = 2\,c_{t\mu}\,h_f\sqrt{\frac{2}{\pi}}\,.
  \end{aligned}
\end{equation}
Setting $D=3$ and using \cref{eq:flow:face}, the integral of
\cref{eq:flow:dupe} evaluates to
\begin{equation}\label{eq:QED:alphaGFfull}
    \begin{aligned}
      \apigfhat&
        =\apiehat\left[1-\apiehat+\frac{\pi}{2}\apiehat^2-\frac{\pi}{4}
          \apiehat^3\,\mathcal{G}\left(\frac{\pi}{4}
          \apiehat^2\right)\right]
    \end{aligned}
\end{equation}
with 
\begin{equation}\label{eq:QED:defF}
    \begin{aligned}
      \mathcal{G}(x)\equiv
      e^{-x}\left[
        \Gamma\left(0,-x\right)+\sqrt{\pi}\,i\Gamma\left(\frac12,-x\right)
        \right]\,,
    \end{aligned}
\end{equation}
where
\begin{equation}\label{eq:QED:hugh}
  \begin{aligned}
    \Gamma(s,z)=\int_z^\infty \text{d}t\ t^{s-1}\ e^{-t}
  \end{aligned}
\end{equation}
is the incomplete $\Gamma$ function. Regulating this integral for $z=-x<0$ by
suitable subtractions, one can write
\begin{equation}\label{eq:QED:eden}
  \begin{aligned}
    \Gamma(0,-x) & = -i\pi + \int_{-x}^1\dd t\,\frac{e^{-t}-1}{t} - \ln x +
    \Gamma(0,1)\,,\\ i\Gamma\left(\frac{1}{2},-x\right)
    &= i\sqrt{\pi} + 2 \sqrt{x} +
    \int_0^x\dd t\,\frac{e^{t}-1}{\sqrt{t}}\,,
  \end{aligned}
\end{equation}
which shows that $\mathcal{G}(x)$ in \cref{eq:QED:defF} is real-valued. Its
large- and small-$x$ limits can be obtained from the usual expansions of the
incomplete $\Gamma$ function.

\begin{figure}
  \begin{center}
    \begin{tabular}{c}
      \raisebox{0em}{%
        \mbox{%
          \includegraphics[%
            clip,width=.6\textwidth]%
                          {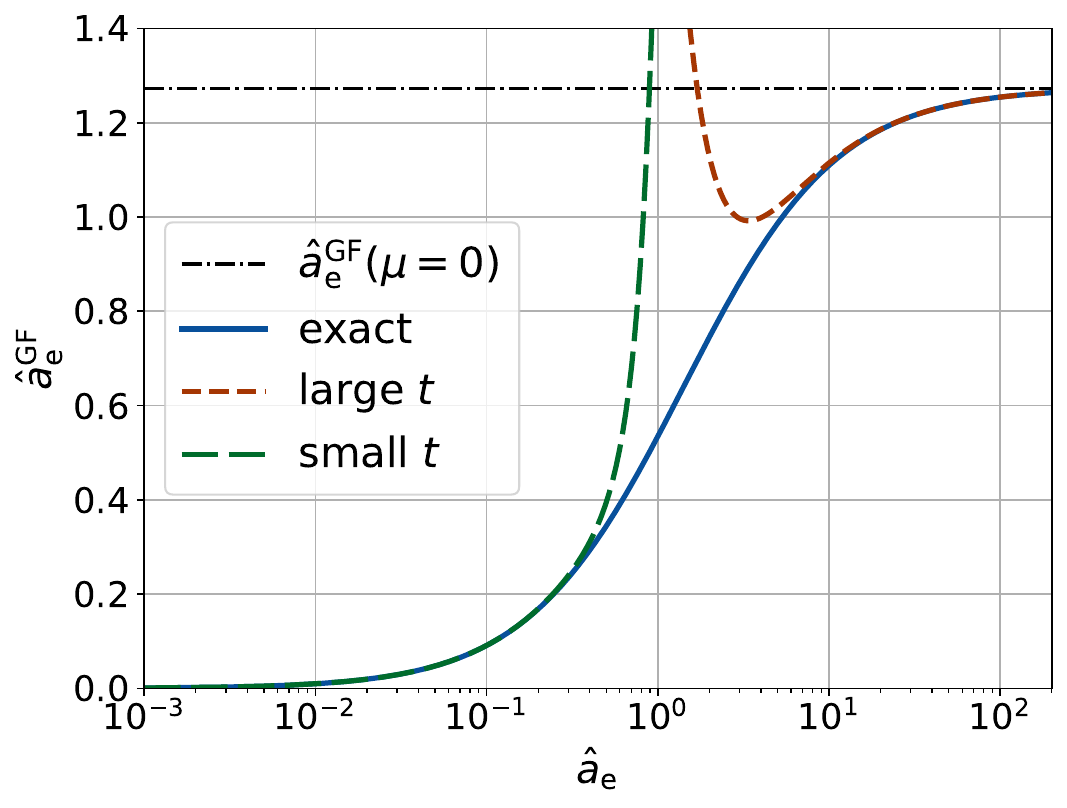}}}
    \end{tabular}
    \parbox{.9\textwidth}{
      \caption[]{\label{fig:alphaQED3}\sloppy The conversion between the GF and $\msbar$ scheme together with the limiting cases for small/large $\apiehat$.}}
  \end{center}
\end{figure}

In this way, we obtain the limits of short and long flow time of
\cref{eq:QED:alphaGFfull}:\footnote{A derivation of these expressions using
the strategy of regions is outlined in \cref{app:sor}.}
\begin{equation}\label{eq:QED:bant}
  \begin{aligned}
    \apigfhat&= \apiehat\left[
      1 - \apiehat + \frac{\pi}{2}\apiehat^2
      + \frac{\pi}{4}\apiehat^3\left(\EulerGamma + 2\ln\apiehat +
      \ln\frac{\pi}{4}\right) + \order{\apiehat^4}\right]\,,\\
      \apigfhat &=
      \frac{4}{\pi}\left[1 - \frac{3}{2\apiehat} + \frac{8}{\pi\apiehat^2}
                           - \frac{15}{\pi\apiehat^3}
        + \order{\apiehat^{-4}}\right]\,.      
  \end{aligned}
\end{equation}
The behaviour of $\apigfhat$ in comparison with these limiting cases is shown
in \cref{fig:alphaQED3}. The
short flow-time limit agrees with the naive expansion of \cref{eq:QED:bain} at
leading order in $1/\nf$.

The dimensionless gradient-flow coupling $\apigfhat$ then clearly has an
\ir\ fixed point at
\begin{equation}\label{eq:QED:iota}
  \begin{aligned}
    \apigfhat(\mu=0) &= \frac{4}{\pi}\,.
  \end{aligned}
\end{equation}
For the following, it is useful to define
\begin{equation}
    \apigfhatstar(\mu) \equiv \apigfhat(\mu) - \frac{4}{\pi}
\end{equation}
with which the inverse relations of \cref{eq:QED:bant} become
\begin{equation}\label{eq:QED:elis}
  \begin{aligned}
    \apiehat
    &= \apigfhat\bigg[
      1 + \apigfhat
      + \left(\apigfhat\right)^2\left(2-\frac{\pi}{2}\right)
      \\&\quad
      + \left(\apigfhat\right)^3
      \left[5-\frac{5\pi}{2} - \frac{\pi}{4}\left(\EulerGamma
        +2\ln\apigfhat + \ln\frac{\pi}{4}\right)
        + \order{(\apigfhat)^4}
        \right]\,,\\
      \frac{1}{\apiehat} &= -\frac{\pi}{6}\apigfhatstar\left[
        1 -\frac{8}{9}\apigfhatstar
        + \frac{256-45\pi^2}{162}(\apigfhatstar)^2 + \cdots\right]\,.
  \end{aligned}
\end{equation}
Using \cref{eq:QED:base,eq:QED:alphaGFfull}, we can obtain
$\beta_{\text{e}}^\text{\gf}$:
\begin{equation}\label{eq:QED:BetafuncComp}
    \begin{aligned}
      \beta_\text{e}^\text{\gf}&=\mu\dderiv{}{}{\mu} \apigfhat
      =-\apiehat \frac{\partial}{\partial \apiehat}\apigfhat\\
      &=-\apiehat\left\{1-2\apiehat+\frac{3\pi}{2}\apiehat^2
      +\frac{\pi}{2}\apiehat^3\left[1-2\,\mathcal{G}
        \left(\frac{\pi}{4}\apiehat^2\right)\right]
      \right.\\
            &\qquad\left.-\frac{\pi^2}{4}\apiehat^4+\frac{\pi^2}{8}\apiehat^5
            \,\mathcal{G}\left(\frac{\pi}{4}\apiehat^2\right)\right\}\,.
    \end{aligned}
\end{equation}
Of course, we are actually interested in $\beta_\text{e}^{\text{\gf}}$ as a
function of $\apigfhat$ rather than $\apiehat$.  However, obtaining an
analytic expression for this would require solving \cref{eq:QED:alphaGFfull}
for $\apiehat\!\left(\apigfhat\right)$, which is beyond analytic reach.
Therefore, the $\beta$ function can only be determined numerically.

\begin{figure}
  \begin{center}
    \begin{tabular}{cc}
      \raisebox{0em}{%
        \mbox{%
          \includegraphics[%
            clip,width=.45\textwidth]%
                          {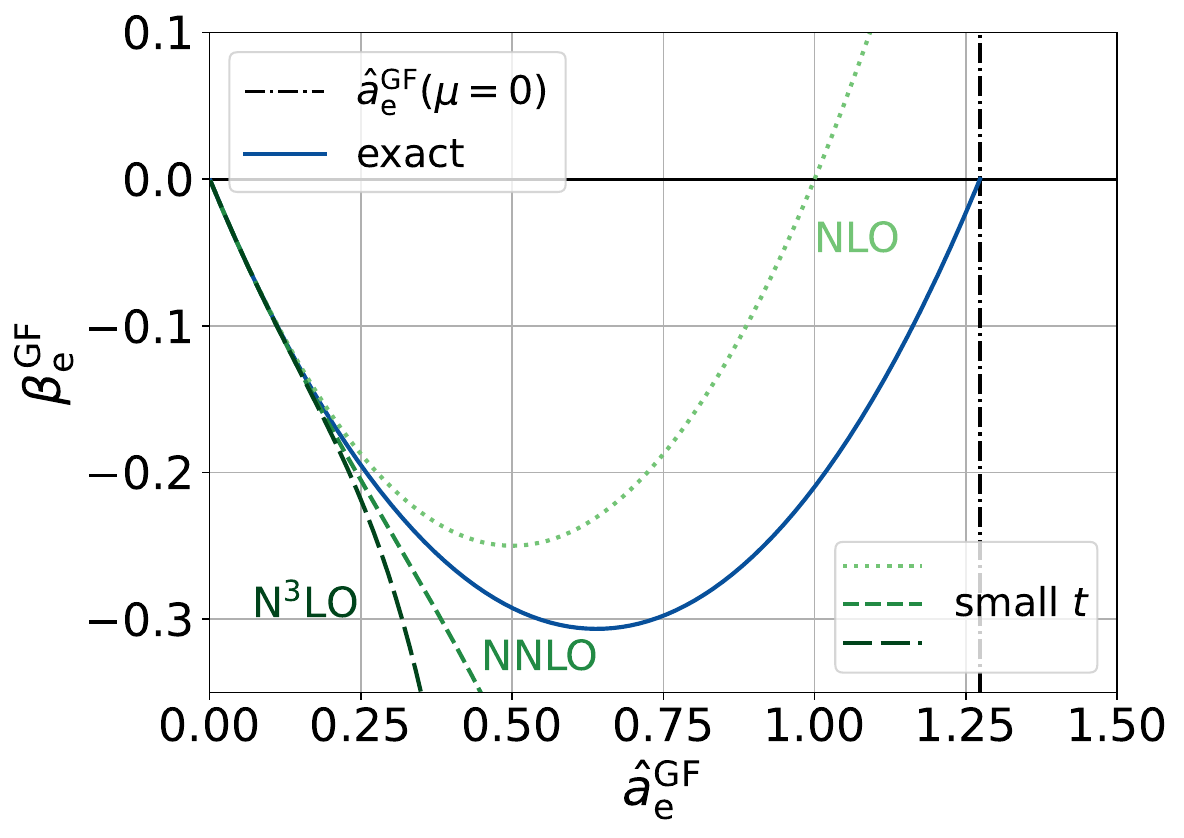}}} &
      \raisebox{0em}{%
        \mbox{%
          \includegraphics[%
            clip,width=.45\textwidth]%
                          {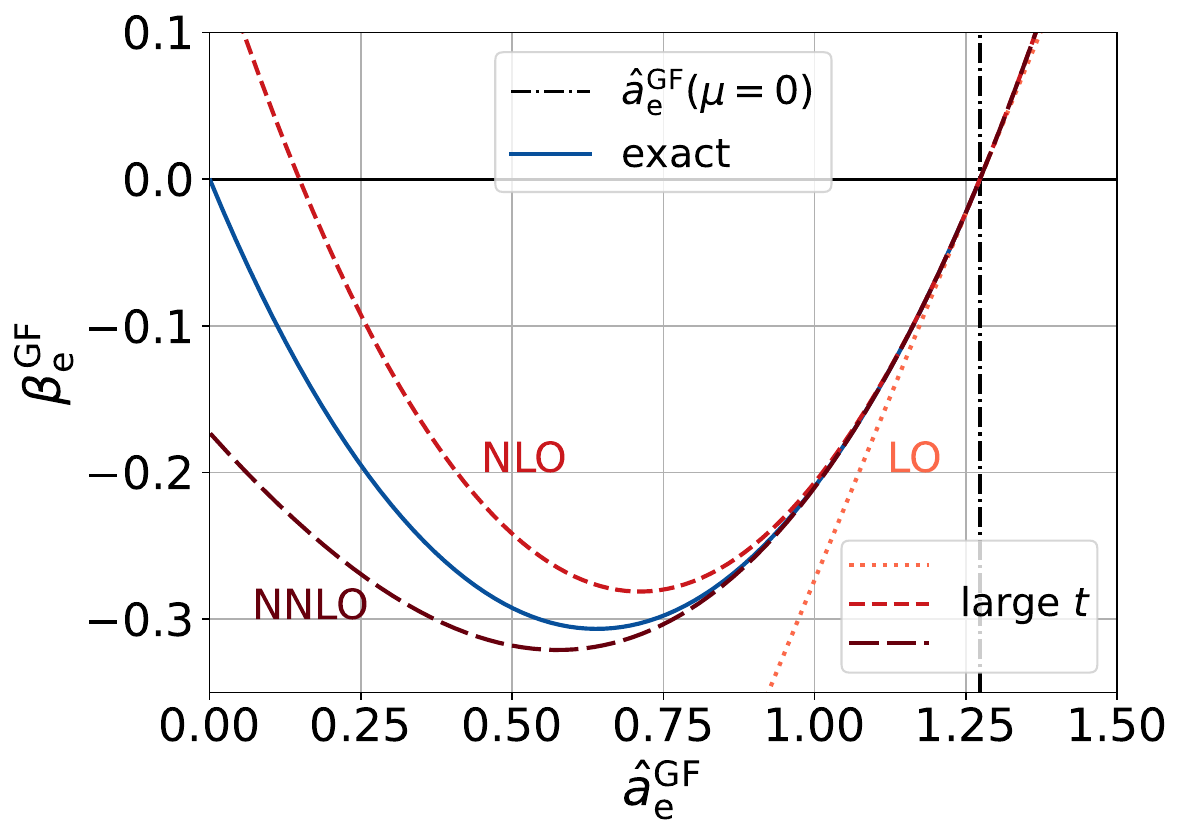}}} \\
      (a) & (b)
    \end{tabular}
    \parbox{.9\textwidth}{
      \caption[]{\label{fig:betaQED3}\sloppy The comparison of the $\beta$ function to the limiting cases: (a) small $\apigfhat$; (b) large $\apigfhat$. } }
  \end{center}
\end{figure}

To gain insight into the behaviour near the fixed points at $\apigfhat=0$ and
$\apigfhat=\frac{4}{\pi}$, we define an approximate $\beta$ function valid in
the limits of short and long flow time.  Expanding \cref{eq:QED:BetafuncComp}
in the respective limits and inserting \cref{eq:QED:elis}, we obtain:
\begin{equation}\label{eq:QED:isai}
  \begin{aligned}
    \mu\dderiv{}{}{\mu} \apigfhat &=
    \beta_\text{e}^\text{\gf}\,,
  \end{aligned}
\end{equation}
with
\begin{equation}\label{eq:QED:juda}
  \begin{aligned}
    \beta_\text{e}^\text{\gf} &= -\apigfhat + (\apigfhat)^2
    +  (\apigfhat)^3(2-\pi)
    \\&\quad
    + (\apigfhat)^4
    \left[5-\frac{3\pi}{2}\left(3+\frac{\EulerGamma}{2}+\ln\apigfhat
      +\frac{1}{2}\ln\frac{\pi}{4}\right)\right] + \cdots
  \end{aligned}
\end{equation}
and
\begin{equation}\label{eq:QED:betaLarge}
    \begin{aligned}
         \beta_\text{e}^\text{\gf}&=\apigfhatstar\left[1+\frac{8}{9}\apigfhatstar-\frac{128-45\pi}{81}(\apigfhatstar)^2+\cdots\right]\,.
    \end{aligned}
\end{equation}
These limits, together with the full $\beta$ function of
\cref{eq:QED:BetafuncComp}, are shown in \cref{fig:betaQED3}.  Note that, for
\cref{eq:QED:juda}, neglecting terms of order $(\apigfhat)^3$ and higher would
imply a non-Gau\ss{}ian fixed point at $\apigfhat=1$. However, we are not
supposed to use this result in the \ir\ limit, which corresponds to
$\sqrt{t}\to \infty$, as is clearly visible in \cref{fig:betaQED3}~(a).

The critical slope, defined as the gradient of the $\beta$-function with
respect to the coupling evaluated at the fixed point, is a scheme independent
critical exponent. Previous studies have found in the large $\nf$ expansion
that, to leading order, $\beta^\prime(a_{\star})=1$
\cite{Palanques-Mestre:1983ogz,Appelquist:1986fd}. By inspection,
\cref{eq:QED:betaLarge} agrees with this result, as does the combination of
the complete analytic results given in \cref{eq:QED:alphaGFfull} and
\cref{eq:QED:BetafuncComp}. $\apigfhat$ approaches the fixed point value as
$\apiehat\to \infty$, and the $\nf$ dependence in these equations is absorbed into our definition of $\apiehat$, therefore we do not expect higher-order
corrections to the gradient-flow estimate of this value, in agreement with the regular perturbative expansion to $\mathcal{O}(1/\nf)$ in
\citere{Palanques-Mestre:1983ogz}, for example. The evaluation of other
critical exponents, such as the fermion-mass anomalous dimension, would
require the computation of flowed fermionic operators, which is outside of the scope of this work.

\section{Conclusions}
\label{sec:conclusions}

We have evaluated the coupling of \four- and \three-dimensional \QED\ with
$\nf$ massless fermions in the gradient-flow scheme. Due to its linear form,
the flow equation can be solved exactly for \QED, and the \gf\ coupling is
fully determined as the Laplace transform of a function that only depends on
the photon polarization function of the unflowed theory. In the latter, one
can include arbitrary additional interactions, so that the U(1) gradient-flow
coupling of a more general theory can be obtained from a \one-dimensional
integral over purely unflowed results within that theory.

As applications of this observation, we have considered \four-dimensional
\QED\ through \four-loop level, as well as \three-dimensional \QED\ in the
large-$\nf$ limit through \nnlo. We have evaluated the corresponding
conversions from the regular to the \gf\ couplings, as well as the associated
$\beta$ functions. For \QEDthree, we recover the well-known \uv\ and
\ir\ fixed points.

We hope that our results will be useful in further field theoretical studies,
in particular in the context of lattice formulations of \QED. 

\paragraph{Acknowledgments.}
We thank Curtis Peterson for bringing our attention to this problem. We also thank him, Nobuyuki Matsumoto and John Gracey for helpful comments on the manuscript, and Jonas Kohnen, Henry Werthenbach and Janosch Borgulat for useful discussions. This research was supported by the Deutsche Forschungsgemeinschaft (DFG, German Research Foundation) under grant 396021762 -- TRR 257.

\begin{appendix}
  
  \section{Long- and short-flow-time limits of
    \boldmath{$E(t)$} in QED\boldmath{$_3$}}\label{app:sor}

In \QEDthree, the integral in
\cref{eq:flow:dupe} depends on the two dimensionful scales $\apie h_f$ and
$t$.  Let us first evaluate it in the limit $\apie h_f\ll
1/\sqrt{t}$. Following the strategy-of-regions~\cite{Beneke:1997zp}, we
identify two integration regions:\footnote{For the application of the
strategy-of-regions to \gf\ integrals, see also \citere{Harlander:2021esn};
for an alternative way to expand \gf\ integrals, see
\citere{Takaura:2025pao}.}
\begin{enumerate}
\item\label{reg:largep} $p\sim 1/\sqrt{t}$: here we can expand the integrand
  in the limit of small $\apie h_f$:
  \begin{equation}\label{eq::john}
    \begin{aligned}
      E^{(1)}(t) &= \frac{e^2(D-1)}{2}
      \int\frac{\dd^D p}{(2\pi)^D}\,e^{-2tp^2}\,
      \sum_{n=0}^\infty\left(-\frac{\apie h_f}{p}\right)^n\\
      &= \frac{e^2(D-1)}{2(8\pi t)^{D/2}\Gamma(D/2)}
      \sum_{n=0}^\infty(-\apie
      h_f\sqrt{2t})^n\, \Gamma\left(\frac{D-n}{2}\right)\,,
    \end{aligned}       
  \end{equation}
  where we have used \cref{eq:flow:blot}. Note that, in the limit $D\equiv
  3-2\ep\to 3$, $E^{(1)}$ has poles for odd $n$ with $n\geq 3$.
\item $p\ll 1/\sqrt{t}$: here we can expand the exponential:\footnote{The
integral was solved with \texttt{Mathematica}~\cite{Mathematica} and using the
identity $\pi/\sin(s\pi) = \Gamma(s)\Gamma(1-s)$~\cite{Miller:book}.}
  \begin{equation}\label{eq::hell}
    \begin{aligned}
      E^{(2)}(t) &= \frac{e^2(D-1)}{(4\pi)^{D/2}\Gamma(D/2)}
      \sum_{n=0}^\infty \frac{1}{n!}
      \int_0^\infty\dd p\,p^{D-1}\frac{(-2p^2t)^n}{1+\apie h_f/p}\\
      &= \frac{e^2(D-1)}{(4\pi)^{D/2}\Gamma(D/2)}
      \sum_{n=0}^\infty \frac{1}{n!} (-2t)^n (\apie
      h_f)^{D+2n}\\&\qquad\quad\times \Gamma(D-1+2n)\Gamma(2(1-n)-D)\,.
    \end{aligned}
  \end{equation}
  Therefore, this region only contributes whenever region~\ref{reg:largep}
  develops poles, in which case they are exactly canceled in the sum of both
  regions. Inserting this sum into \cref{eq:flow:bund} results in the first
  equation of \cref{eq:QED:bant}.
\end{enumerate}

In the opposite limit, $\apie h_f\gg 1/\sqrt{t}$, the region $p\gg 1/\sqrt{t}$
vanishes due to the exponential. Therefore, only the region $p\sim 1/\sqrt{t}$
contributes, leading to
\begin{equation}\label{eq::boat}
  \begin{aligned}
    E(t) &= -\frac{e^2(D-1)}{2}
      \int\frac{\dd^D p}{(2\pi)^D}\,e^{-2tp^2}\,
      \sum_{n=0}^\infty\left(-\frac{p}{\apie h_f}\right)^{n+1}\\
      &= -\frac{e^2(D-1)}{2(8\pi t)^{D/2}\Gamma(D/2)}\sum_{n=0}^\infty
      \left(-\frac{1}{\apie h_f \sqrt{2t}}\right)^{n+1}
          \Gamma\left(\frac{D+n+1}{2}\right)\,.
  \end{aligned}
\end{equation}
Inserting this into \cref{eq:flow:bund} results in the second equation of
\cref{eq:QED:bant}.

\section{QED\boldmath{$_4$} gradient-flow coupling including QCD corrections}\label{app:qcd}

Here we provide the explicit result for the \qcd\ corrections to the
\QEDfour\ coupling in the \gf\ scheme. As a consequence of
\cref{eq::does}, we can write
\begin{equation}\label{eq::aria}
  \begin{aligned}
    \apigf(\mu) &= \apie(\mu)\left\{1+\apie(\mu) e_1 +
    \apie(\mu)\left[\sum_{n\geq
        1}e_n^\text{\qcd}\apis^n(\mu)\right]\right\}\,,
  \end{aligned}
\end{equation}
where the coefficients $e_n^{\text{\qcd}} = e_n^{\text{\qcd}}(\lmut)$ can be extracted using $d_{n,m}$ from Ref.~\cite{Baikov:2012zm} and, 
up to $n = 2$, take the form
\begin{equation}\label{eq::kail}
  \begin{aligned}
    e_1^\text{\qcd}&=-d_{1,0}+d_{1,1}\left(1-\lmut\right)=\left(\zeta(3)-\frac{43}{48}-\frac{1}{4}\lmut\right)\dR\nf\ccf\,,\\
    e_2^\text{\qcd}&=-d_{2,0}+d_{2,1}(1-\lmut)+d_{2,2}\left(-\frac{\pi^2}{6}+2\lmut-\lmut^2\right)\\
    &=\dR\nf\ccf\cca\left[\frac{5}{12}\zeta(5)+\frac{161}{72}\zeta(3)-\frac{11\pi^2}{576}-\frac{30931}{10368}+\lmut\left(\frac{11}{12}\zeta(3)-\frac{101}{96}\right)-\frac{11}{96}\lmut^2\right]\\
    &\quad+\dR\nf\ccf^2\left(-\frac{5}{2}\zeta (5)+\frac{37}{24}\zeta (3)+\frac{67}{144}+ \frac{1}{32}\lmut
\right)\\
&\quad+\dR\nf^2\ccf\ctr\left[-\frac{13 }{18}\zeta (3)+\frac{\pi ^2}{144}+\frac{2513}{2592}-\lmut\left(\frac{1}{3} \zeta (3)-\frac{3}{8}\right)+\frac{1}{24}\lmut^2
\right]\,.
\end{aligned}
\end{equation}
The coefficient $e_1$ coincides with the one in \QEDfour,
cf.\ \cref{eq::en}. The relevant SU$(3)$ group factors required for \qcd\ are:
\begin{equation}
    \begin{aligned}
        \dR=3\,, \quad \ccf=\frac{4}{3}\,,\quad\cca=3\,,\quad \ctr=\frac{1}{2}\,.
    \end{aligned}
\end{equation}
The derivation of higher-order terms in \cref{eq::aria} proceeds analogously.

\end{appendix}

\bibliography{literatur}

\end{document}